\documentclass[useAMS,usenatbib]{mn2e}
\usepackage{epsfig}
\usepackage{float}
\usepackage{placeins}
\usepackage{graphicx}
\usepackage{epsfig}
\usepackage{bm}
\usepackage{amsfonts}
\usepackage{amssymb}
\usepackage{times}
\usepackage{natbib}
\usepackage{color}

\title[Optical variability of OJ 287]
{Multiband optical variability of the blazar OJ 287 during its outbursts in 2015 -- 2016}

\author[Gupta et al.]
{Alok C.\ Gupta$^{1,2}$\thanks{Email: acgupta30@gmail.com}\thanks{CAS PIFI Visiting Scientist},
Aditi Agarwal$^{2}$\thanks{Email:aditi@aries.res.in},
Alka Mishra$^{2}$, 
H. Gaur$^{1}$,
P. J. Wiita$^{3}$,
M. F. Gu$^{1}$,
\newauthor O. M. Kurtanidze$^{4,5}$,
G. Damljanovic$^{6}$,
M. Uemura$^{7}$,
E. Semkov$^{8}$,
A. Strigachev$^{8}$,
\newauthor R. Bachev$^{8}$,
O. Vince$^{6}$,
Z. Zhang$^{1}$,
B. Villarroel$^{9}$,
P. Kushwaha$^{10}$,
A. Pandey$^{2}$,
T. Abe$^{7}$,
\newauthor R. Chanishvili$^{4}$, 
R. A. Chigladze$^{4}$,
J. H. Fan$^{11}$,
J. Hirochi$^{7}$,
R. Itoh$^{12}$,
Y. Kanda$^{7}$,
\newauthor M. Kawabata$^{7}$, 
G. N. Kimeridze$^{4}$,
S. O. Kurtanidze$^{4}$,
G. Latev$^{8}$,
R. V. Mu\~{n}oz Dimitrova$^{8}$,
\newauthor T. Nakaoka$^{7}$,
M. G. Nikolashvili$^{4}$,
K. Shiki$^{7}$,
L. A. Sigua$^{4}$,
B. Spassov$^{8}$
\\
\\
\\
$^1$Shanghai Astronomical Observatory, Chinese Academy of Sciences, 80 Nandan Road, Shanghai 200030, China \\
$^2$Aryabhatta Research Institute of Observational Sciences (ARIES), Manora Peak, Nainital, 263 002, India \\
$^3$Department of Physics, The College of New Jersey, P.O. Box 7718, Ewing, NJ 08628-0718, USA \\
$^4$Abastumani Observatory, Mt. Kanobili, 0301 Abastumani, Georgia \\
$^5$Engelhardt Astronomical Observatory, Kazan Federal University, 420 000 Tatarstan, Russia \\
$^6$Astronomical Observatory, Volgina 7, 11060 Belgrade, Serbia \\
$^7$Hiroshima Astrophysical Science Center, Hiroshima University, Kagamiyama 1-3-1, Higashi-Hiroshima 739-8526, Japan \\
$^8$Institute of Astronomy and National Astronomical Observatory, Bulgarian Academy of Sciences, 72 Tsarigradsko Shosse Blvd., 1784 Sofia, Bulgaria \\
$^{9}$Department of Physics and Astronomy, Uppsala Universitet, Box 516, 751 20, Uppsala, Sweden \\
$^{10}$Department of Astronomy (IAG-USP), University of Sao Paulo, Sao Paulo 05508-900, Brazil \\
$^{11}$Center for Astrophysics, Guangzhou University, Guangzhou 510006, China \\
$^{12}$Department of Physics, Tokyo Institute of Technology, 2-12-1 Ookayama, Meguro-ku, Tokyo 152-8551, Japan} 

\begin{document}
\date{Accepted \dots Received \dots; in original form \dots}

\maketitle

\label{firstpage}

\begin{abstract}
We present recent optical photometric observations of the blazar OJ 287 taken during September 2015 -- May 2016. 
Our intense observations of the blazar started in November 2015 and continued until May 2016 and included detection 
of the large optical outburst in December 2016 that was predicted using the binary black hole model for OJ 287. For 
our observing campaign, we used a total of 9 ground based optical telescopes of which one is in Japan, one is in India, 
three are in Bulgaria, one is in Serbia, one is in Georgia, and two are in the USA. These observations were carried 
out in 102 nights with a total of $\sim$ 1000 image frames in BVRI bands, though the majority were in the R band. We 
detected a second comparably strong flare in March 2016.  In addition, we investigated multi-band flux variations, 
colour variations, and spectral changes in the blazar on diverse timescales as they are useful in understanding the 
emission mechanisms. We briefly discuss the possible physical mechanisms most likely responsible for the observed flux, 
colour and spectral variability.
\end{abstract}


\section{Introduction}

Blazars are  a subclass of radio loud active galactic nuclei and are further comprised of two classes, BL Lacertae objects, 
which have either very weak (EW $< 5$\AA ) or no emission lines (Stocke et al.\ 1991; Marcha et al.\ 1996), and flat spectrum 
radio quasars with strong emission lines (e.g., Blandford \& Rees 1978; Ghisellini et al.\ 1997). Blazars are primarily 
characterized by high brightness, high polarization in radio to optical bands, and  highly variable, predominantly
non-thermal emission across the entire electromagnetic (EM) spectrum. The emission is normally attributed to the relativistic 
jet aligned at a small angle with observer's line of sight (LOS; e.g., Urry \& Padovani 1995). The blazar emission over the 
entire EM spectrum reveals the presence of a broad double peaked structures in their spectral energy distributions (SEDs).  
The presence of high polarization and its variation on timescales of the flux variation implies that the low energy peak 
(with usually between the IR and soft X-rays) of the blazar SED is result of synchrotron emission from relativistic non-thermal
electrons in the jet.  The high energy component, which peaks in gamma-rays somewhere between GeV and TeV energies  probably 
originates from inverse Compton up-scattering of synchrotron or external photons off the relativistic electrons producing the 
synchrotron emission (Kirk, Rieger \& Mastichiadis 1998; Gaur et al. 2010).
    
Blazar flux variations over the complete EM spectrum reveal diverse timescales with variability timescales extending from 
a few minutes to years and even decades. Flux variation from  minutes to less than a day are commonly known as intraday 
variability (IDV) (Wagner \& Witzel 1995) or intra-night variability or micro-variability (Goyal et al.\ 2012) while that 
from  days to a few months is often called short term variability (STV) while flux variation over timescale of several months 
to years is usually called long term variability (LTV; Gupta et al.\ 2004). For most blazars the LTV and much of the STV can 
be explained through the shock-in-jet model (e.g. Marscher \& Gear 1985; Hughes, Aller \& Aller 1985). The most puzzling 
variability of blazars is that detected on IDV timescales, and understanding this may allow for probing the very inner
region close the central black hole and might be helpful in measuring the mass of the central super massive black hole mass 
of blazars (Gupta et al. 2012). The first convincing optical flux variability on IDV timescale was reported by Miller et al.\ 
(1989) and since then it has been extensively studied by various groups around the globe (e.g. Carini et al.\ 1991, 1992; 
Carini \& Miller 1992; Heidt \& Wagner 1996; Sagar et al. 1999; Wu et al.\ 2005, 2007; Montagni et al.\ 2006; Stalin et al.\ 2006; 
Gupta et al.\ 2008, 2016; Gaur et al.\ 2012a, 2012b, 2012c; Agarwal \& Gupta 2015; Agarwal et al.\ 2015, 2016; and references 
therein). Optical variability of blazars on STV timescales are useful in exploring flux, colour and spectral variations and 
can also help to separate out any quasi-thermal (e.g. accretion disc emission) from the nonthermal emission component. We 
have done extensive work in this area  (e.g.\ Gu et al.\ 2006; Fan et al. 1998, 2009; Gupta et al. 2008, 2016; Gaur et al. 2012b, 
2012c; Agarwal et al.\ 2015, 2016 and references therein). 

OJ 287 ($\alpha_{2000.0} =$ 08h 54m 48.87s, $\delta_{2000.0} =$ +20$^{\circ}$ 06$^{'}$ 30.$^{''}$64) is a blazar at  redshift 
$z =$ 0.306 (Sitko \& Junkkarinen 1985) that has relatively low frequency peaks for its SED. Its relative brightness means 
that it is one of the best observed blazars with more than a century of observations  in the optical bands. Using this optical
light curve data of OJ 287 starting in 1890, Sillanp{\"{a}}{\"{a}} et al.\ (1988) noticed for the first time that there is 
double-peaked outburst feature which repeated with a period of $\sim$ 12 yrs. To explain this unique quasi-periodic light curve 
feature Sillanp{\"{a}}{\"{a}} et al. (1988) proposed a binary black hole system for this blazar and predicted next double-peaked 
outburst would occur in 1994 -- 1995; this was indeed observed. In this case the second peak, occurring $\sim$ 1.2 years after the detection 
of the first peak, was seen Sillanp{\"{a}}{\"{a}} et al.\ (1996a, 1996b).

A double peaked outburst was also seen during the next recurrence in 2005 -- 2007 with the outburst's first peak found at the 
end of 2005 and the second peak seen at the end of 2007 (Valtonen et al.\ 2009). These predicted recurrences are extremely strong 
evidence for OJ 287 housing a binary black hole system with a decaying orbital period of $\sim$ 12 years. A puzzling issue in 
this model is the timing and strength of the second major peak of a outburst. In the first planned campaign in 1994 -- 1995, the 
second peak was detected  $\sim$ 1.2 years after the first peak (Sillanp{\"{a}}{\"{a}}  et al.\ 1996b) while it was observed  
$\sim$ 2.0 years after the  1st peak during the 2005 -- 2008 campaign (Valtonen et al.\ 2009).

The recurring outbursts of this source are not quite periodic and the quasi-periodic pattern of optical outbursts
was explained by a model where a secondary black hole in a $\sim$ 12 year orbit impacts the accretion
disk of the primary black hole as it orbits the primary (Lehto \& Valtonen 1996). The quasi-Keplerian
nature of binary black hole orbits in general relativity mean that the times of the first impact flare and their associated 
radiation can be predicted (Memmesheimer et al.\ 2004). The deviations from strict
periodicity are understood in a model that contains a gravitational wave driven inspiraling spinning
binary black hole system at its center (Valtonen et al.\ 2008; 2010). In this model, the impact outbursts
are generated by expanding bubbles of hot gas that have been shocked and pulled out of the accretion disk
(Lehto \& Valtonen 1996; Pihajoki 2016). Hence, thermal radiation emanates from the vicinity of the impact
site. Recently, the predicted first peak of the outburst was detected in December 2015 which is the brightest optical
level in 30 years (Valtonen et al.\ 2016) and shows a clear signature of thermal emission.  Valtonen et al.\ (2016)
show that the data collected over the past three cycles give good measurements for the black hole masses,
with the primary at $(1.83\pm 0.01) \times 10^{10}$  and the secondary at $(1.5\pm 0.1) \times 10^{8}$ solar masses.
Remarkably, the spin of the primary can also be determined to high precision, at $a = 0.313 \pm 0.01$.

In the present work, we report detailed optical measurements of the late 2015 outburst OJ 287, including  the detection of an a second
 outburst that is essentially as bright as the 
first one but occurred only  $\sim$ 3 months later. 
The paper is structured as follows: in Section 2, we discuss our new optical observations and data reduction,
as well as public archival data used here. Section 3 gives information about the various techniques
used in the work. In Section 4 we present results and give a discussion of them in Section 5.
Our conclusions are summarized in Section 6.           

\begin{table*}
\caption{Observation log of optical photometric observations of the blazar OJ 287. }
\textwidth=6.0in
\textheight=10.0in
\noindent
\begin{tabular}{lcclcclcc} \hline
~~~~Date & Telescope  & Data Points & ~~~~Date & Telescope  & Data Points &  ~~~~Date & Telescope  & Data Points  \\
yyyy mm dd     &            &B, V, R, I & yyyy mm dd     &            &B, V, R, I & yyyy mm dd     &            &B, V, R, I  \\\hline
2015 09 20  & A & 00, 01, 01, 00         & 2016 01 12  & A & 00, 01, 01, 00      & 2016 03 12  & A & 00, 01, 01, 00 \\
2015 10 13  & A & 00, 01, 01, 00         & 2016 01 12  & F & 01, 01, 01, 01      & 2016 03 13  & A & 00, 02, 02, 00 \\
2015 10 14  & A & 00, 01, 01, 00         & 2016 01 13  & A & 00, 01, 01, 00      & 2016 03 14  & A & 00, 01, 01, 00 \\
2015 10 15  & A & 00, 01, 01, 00         & 2016 01 14  & A & 00, 01, 01, 00      & 2016 03 15  & A & 00, 02, 02, 00 \\
2015 10 29  & G & 00, 00, 04, 00         & 2016 01 14  & H & 00, 01, 01, 00      & 2016 03 16  & A & 00, 02, 02, 00 \\
2015 11 10  & A & 00, 01, 01, 00         & 2016 01 15  & A & 00, 01, 01, 00      & 2016 03 16  & G & 00, 00, 04, 00 \\
2015 11 11  & A & 00, 01, 01, 00         & 2016 01 16  & A & 00, 01, 01, 00      & 2016 03 17  & A & 00, 02, 02, 00 \\
2015 11 12  & A & 00, 01, 01, 00         & 2016 01 18  & A & 00, 01, 01, 00      & 2016 03 30  & G & 00, 00, 08, 00 \\
2015 11 12  & F & 01, 01, 01, 01         & 2016 01 19  & A & 00, 01, 01, 00      & 2016 03 31  & G & 00, 00, 09, 00 \\
2015 11 13  & A & 00, 01, 01, 00         & 2016 01 20  & G & 00, 00, 04, 00      & 2016 04 01  & G & 00, 00, 04, 00 \\
2015 11 13  & G & 00, 00, 03, 00         & 2016 01 30  & H & 00, 01, 01, 00      & 2016 04 03  & G & 00, 00, 04, 00 \\
2015 11 14  & A & 00, 01, 01, 00         & 2016 02 03  & G & 00, 00, 05, 00      & 2016 04 05  & E & 02, 02, 02, 02 \\
2015 11 18  & A & 00, 01, 01, 00         & 2016 02 04  & G & 00, 00, 04, 00      & 2016 04 06  & E & 02, 02, 02, 02 \\
2015 12 04  & F & 01, 01, 01, 01         & 2016 02 05  & G & 00, 00, 04, 00      & 2016 04 08  & B & 01, 06, 06, 01 \\
2015 12 05  & F & 01, 01, 01, 01         & 2016 02 06  & G & 00, 00, 06, 00      & 2016 04 10  & B & 01, 03, 03, 01 \\
2015 12 06  & F & 01, 01, 01, 01         & 2016 02 06  & F & 01, 01, 01, 01      & 2016 04 11  & G & 00, 00, 07, 00 \\
2015 12 07  & G & 00, 00, 17, 00         & 2016 02 07  & H & 00, 01, 01, 00      & 2016 04 11  & B & 01, 23, 23, 01 \\
2015 12 08  & G & 00, 00, 04, 00         & 2016 02 07  & F & 01, 01, 01, 01      & 2016 04 12  & B & 01, 24, 24, 01 \\
2015 12 08  & F & 01, 01, 01, 01         & 2016 02 09  & A & 00, 01, 01, 00      & 2016 04 13  & B & 01, 20, 20, 01 \\
2015 12 09  & A & 00, 02, 02, 00         & 2016 02 10  & A & 00, 01, 01, 00      & 2016 04 13  & A & 00, 01, 01, 00 \\
2015 12 09  & G & 00, 00, 08, 00         & 2016 02 11  & A & 00, 01, 01, 00      & 2016 04 14  & A & 00, 01, 01, 00 \\
2015 12 10  & A & 00, 01, 01, 00         & 2016 02 11  & H & 00, 01, 01, 00      & 2016 04 14  & H & 00, 01, 01, 00 \\
2015 12 10  & G & 00, 00, 14, 00         & 2016 02 12  & A & 00, 03, 03, 00      & 2016 04 16  & G & 00, 00, 06, 00 \\
2015 12 11  & G & 00, 00, 09, 00         & 2016 02 13  & A & 00, 01, 01, 00      & 2016 04 18  & H & 00, 01, 01, 00 \\
2015 12 12  & G & 00, 00, 11, 00         & 2016 02 14  & A & 00, 01, 01, 00      & 2016 04 26  & E & 03, 03, 03, 03 \\
2015 12 12  & F & 01, 01, 01, 01         & 2016 02 15  & A & 00, 01, 01, 00      & 2016 04 27  & E & 09, 09, 09, 09 \\
2015 12 12  & C & 01, 01, 04, 01         & 2016 02 15  & G & 00, 00, 18, 00      & 2016 04 27  & D & 01, 03, 03, 01 \\
2015 12 12  & E & 02, 00, 02, 02         & 2016 02 16  & A & 00, 03, 03, 00      & 2016 04 29  & B & 01, 03, 03, 01 \\
2015 12 13  & G & 00, 00, 07, 00         & 2016 02 17  & H & 00, 01, 01, 00      & 2016 05 01  & G & 00, 00, 11, 00 \\
2015 12 13  & F & 01, 01, 01, 01         & 2016 02 17  & G & 00, 00, 06, 00      & 2016 05 03  & A & 00, 01, 01, 00 \\
2015 12 13  & C & 01, 01, 04, 01         & 2016 02 25  & H & 00, 01, 01, 00      & 2016 05 03  & H & 00, 01, 01, 00 \\
2015 12 13  & E & 02, 00, 02, 02         & 2016 02 29  & H & 00, 01, 01, 00      & 2016 05 04  & A & 00, 01, 01, 00 \\
2015 12 14  & A & 00, 02, 02, 00         & 2016 02 29  & G & 00, 00, 04, 00      & 2016 05 05  & A & 00, 01, 01, 00 \\
2015 12 14  & F & 01, 01, 01, 01         & 2016 03 01  & G & 00, 00, 04, 00      & 2016 05 06  & A & 00, 01, 01, 00 \\
2015 12 14  & C & 01, 01, 06, 01         & 2016 03 01  & E & 02, 02, 02, 02      & 2016 05 06  & H & 00, 01, 00, 00 \\
2015 12 14  & E & 02, 00, 02, 02         & 2016 03 02  & G & 00, 00, 06, 00      & 2016 05 08  & A & 00, 01, 01, 00 \\
2015 12 15  & F & 01, 01, 01, 01         & 2016 03 02  & E & 02, 02, 02, 02      & 2016 05 09  & A & 00, 01, 01, 00 \\
2015 12 15  & D & 01, 01, 06, 01         & 2016 03 03  & G & 00, 00, 04, 00      & 2016 05 10  & A & 00, 01, 01, 00 \\
2015 12 16  & A & 00, 02, 02, 00         & 2016 03 04  & G & 00, 00, 02, 00      & 2016 05 11  & A & 00, 01, 01, 00 \\
2015 12 17  & D & 01, 01, 04, 01         & 2016 03 05  & G & 00, 00, 05, 00      & 2016 05 11  & H & 00, 01, 01, 00 \\
2015 12 18  & G & 00, 00, 35, 00         & 2016 03 06  & G & 00, 00, 02, 00      & 2016 05 13  & D & 01, 03, 03, 01 \\
2015 12 19  & G & 00, 00, 33, 00         & 2016 03 06  & F & 01, 01, 01, 01      & 2016 05 14  & D & 01, 03, 03, 01 \\
2015 12 21  & G & 00, 00, 06, 00         & 2016 03 07  & H & 00, 01, 00, 00      & 2016 05 17  & H & 00, 01, 01, 00 \\
2015 12 26  & G & 00, 00, 08, 00         & 2016 03 09  & A & 00, 01, 01, 00      & 2016 05 20  & H & 00, 01, 01, 00 \\
2015 12 26  & H & 00, 00, 01, 00         & 2016 03 10  & A & 00, 02, 02, 00      & 2016 05 31  & C & 02, 02, 02, 02 \\
2016 01 06  & H & 00, 01, 01, 00         & 2016 03 11  & A & 00, 02, 02, 00      &                 &   &   \\
2016 01 10  & H & 00, 01, 01, 00         & 2016 03 11  & H & 00, 01, 01, 00      &                 &   &   \\\hline
\end{tabular} \\
A: 2.3-m Bok Telescope and 1.54-m Kuiper Telescope at Steward Observatory, Arizona, USA \\
B: 1.04-m Sampuranand Telescope, ARIES, Nainital, India.  \\
C: 2-m Ritchey-Chretien telescope at National Astronomical Observatory, Rozhen, Bulgaria. \\
D: 50/70-cm Schmidt telescope at National Astronomical Observatory, Rozhen, Bulgaria. \\
E: 60-cm Cassegrain telescope at Astronomical Observatory, Belogradchik, Bulgaria. \\
F: 60-cm Cassegrain telescope, Astronomical Station Vidojevica (ASV), Serbia \\
G: 70-cm  meniscus telescope at Abastumani Observatory, Georgia \\
H: 1.5-m KANATA telescope at Higashi–Hiroshima Observatory, Japan \\
\end{table*}

\section{Observations and Data Reduction}

\subsection{New Optical Observations and Data Reduction} 

Our optical photometric observing campaign of the blazar OJ 287 in B, V, R, and I passbands were
carried out using telescopes in India,  Bulgaria (3),  Serbia,  Georgia and Japan and correspond to
telescopes $B$ through $H$ in the notes to Table 1.
These telescopes are equipped with CCD detectors and broad band optical filters B, V, R, and I. Details of
the  CCDs mounted on telescopes B through F are reported in our earlier papers (Agarwal et al.\ 2015; 
Gupta et al.\ 2016). Information about telescopes G and H and their CCDs are given in Table 2. Telescope
A corresponds to archival data from Steward Observatory.

We have performed multiband optical observations of the blazar OJ 287 (see Table 1) spanning
the  period between September 2015 and May 2016. These photometric observations include
R band monitoring as continuously as we could manage so as  to explore IDV.  Quasi-simultaneous monitoring in 
BVRI filters were frequently made
to study the day to day variations in the brightness and colour of the blazar. For the five telescopes $B$ to $F$
bias frames were taken at regular intervals during each night. Twilight sky flat-field frames were
captured to get uniform light in the dawn and/or dusk of each observational night in all the filters.
Observations were taken in the 1 $\times$ 1, 2 $\times$ 2 and 3 $\times$ 3 binning mode on different
telescopes to improve signal to noise (S/N) ratio as discussed earlier  (Agarwal et al.\ 2015; 
Gupta et al.\ 2016).   Image preprocessing involved the removal of detector
effects through bias subtraction, flat fielding, edge trimming and cosmic ray removal which we did using
 standard procedures in  IRAF\footnote{IRAF is distributed by the National Optical Astronomy
Observatories, which are operated by the Association of Universities for Research in Astronomy, Inc.,
under cooperative agreement with the National Science Foundation.}. After the preprocessing we carry out the data reduction,
for which standard photometry software, the Dominion Astronomical Observatory
Photometry (DAOPHOT II) (Stetson 1987; Stetson 1992), was used to obtain 
instrumental magnitude of the target and  comparison stars. Since blazars fields are not crowded  and
we are interested only in the blazar and the comparison stars which are of comparable brightness, aperture
photometry is sufficient to get their instrumental magnitudes. Other than our source, we also observed
more than three local standard stars (e.g. Smith et al.\ 1985; Fiorucci \& Tosti 1996; Gonz{\'a}lez-P{\'e}rez et al.\ 2001)
in the same field of view to calibrate magnitude of the blazar and also to check for the presence of
variability. Detailed descriptions of the  data analysis are given in our earlier papers 
(e.g.\ Gaur et al.\ 2012c; Agarwal \& Gupta 2015; Agarwal et al.\ 2015; Gupta et al.\ 2016; and references therein). 

\begin{table}
\caption{Details of telescopes and instruments}
\hspace*{-0.2in}
\noindent
\begin{tabular}{lll} \hline
                & Abastumani, Georgia    & KANATA, Japan \\\hline
Telescope       & 70-cm Meniscus         & 1.5-m Ritchey-Chretian \\
CCD Model       & Apogee Ap6E (KAF-1001) & Hamamatsu fully-depleted  \\
Chip Size       & 1024 $\times$ 1024 pixels$^{2}$ & 2048 $\times$ 2048 pixels$^{2}$ \\
Scale           & 2.4 arc sec pixel$^{-1}$ & 0.29 arc sec pixel$^{-1}$ \\
Field           & 15 $\times$ 15 arcmin$^{2}$ & 10 $\times$ 10 arcmin$^{2}$ \\
Gain            & 8 e$^{-1}$ ADU$^{-1}$ & 2.2 e$^{-1}$ ADU$^{-1}$ \\
Read out noise  & 14 e$^{-1}$ rms       & 5 e$^{-1}$ rms \\
Typical Seeing  & 1 -- 2 arcsec         & 1 -- 2 arcsec \\\hline
\end{tabular}
\end{table} 

Observations of OJ 287 at the Abastumani Observatory were conducted from 2015 October 29 to 2016 May 1 at
the 70 cm meniscus telescope (f/3) (telescope $G$ hereafter). These measurements were made with an
Apogee CCD camera Ap6E (1K $\times$ 1K, 24 $\mu$m$^{2}$ pixels) through a Cousins R filter. Reduction
of the image frames was done using DAOPHOT II. An aperture radius of 5 arcsec was used for data analysis.
The blazar data is calibrated using the local standard star number 4 in the field (Fiorucci $\&$ Tosti 1996).

We performed  V and R band photometric observations of the blazar OJ 287 using the HONIR instrument
installed on the 1.5 m Kanata telescope (telescope $H$ hereafter) in the Higashi-Hiroshima Observatory,
Japan (Akitaya et al.\ 2014). The data were reduced using standard procedures for CCD photometry. We performed
aperture photometry using the APPHOT package in IRAF, and the blazar data is calibrated with the local 
standard star number 10 (Fiorucci $\&$ Tosti 1996). 

\subsection{Archival Optical Data} 

We also include recent optical photometric data taken from an archive of observations made at Steward Observatory, University of 
Arizona, USA\footnote{http://james.as.arizona.edu/~psmith/Fermi/datause.html}. The observations are carried
out using the 2.3m Bok and 1.54m Kupier telescopes. These are clubbed together  as telescope $A$ in the observation log 
presented in Table 1. These photometric observations of the blazar OJ 287 were made using the SPOL CCD 
Imaging/Spectropolarimeter mounted on those two telescopes. Details about the instrument, observation and data 
analysis are described in Smith et al.\ (2009).

\begin{figure*}
\epsfig{figure= 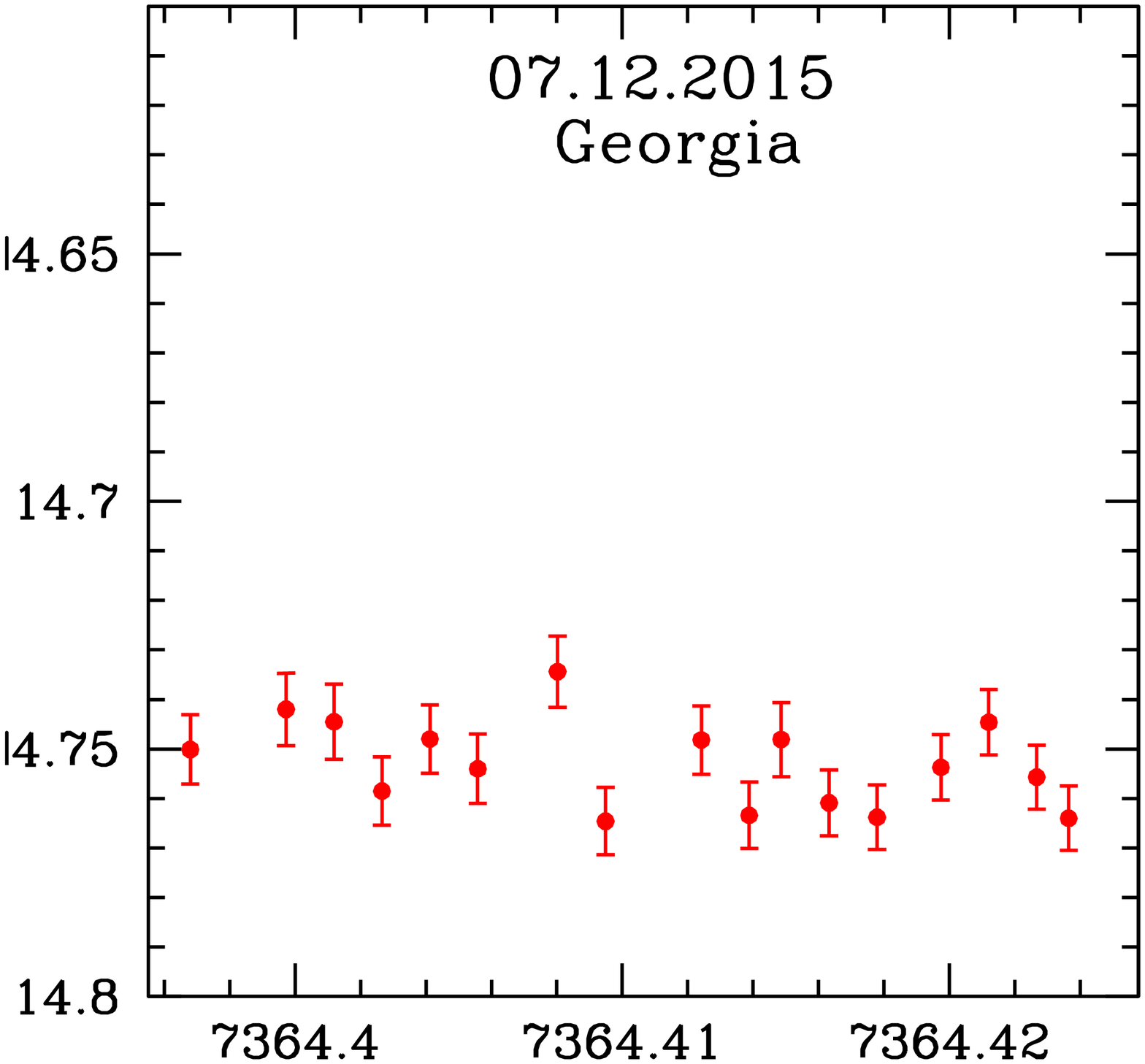,height=2.0in,width=2.0in,angle=0}
\epsfig{figure= 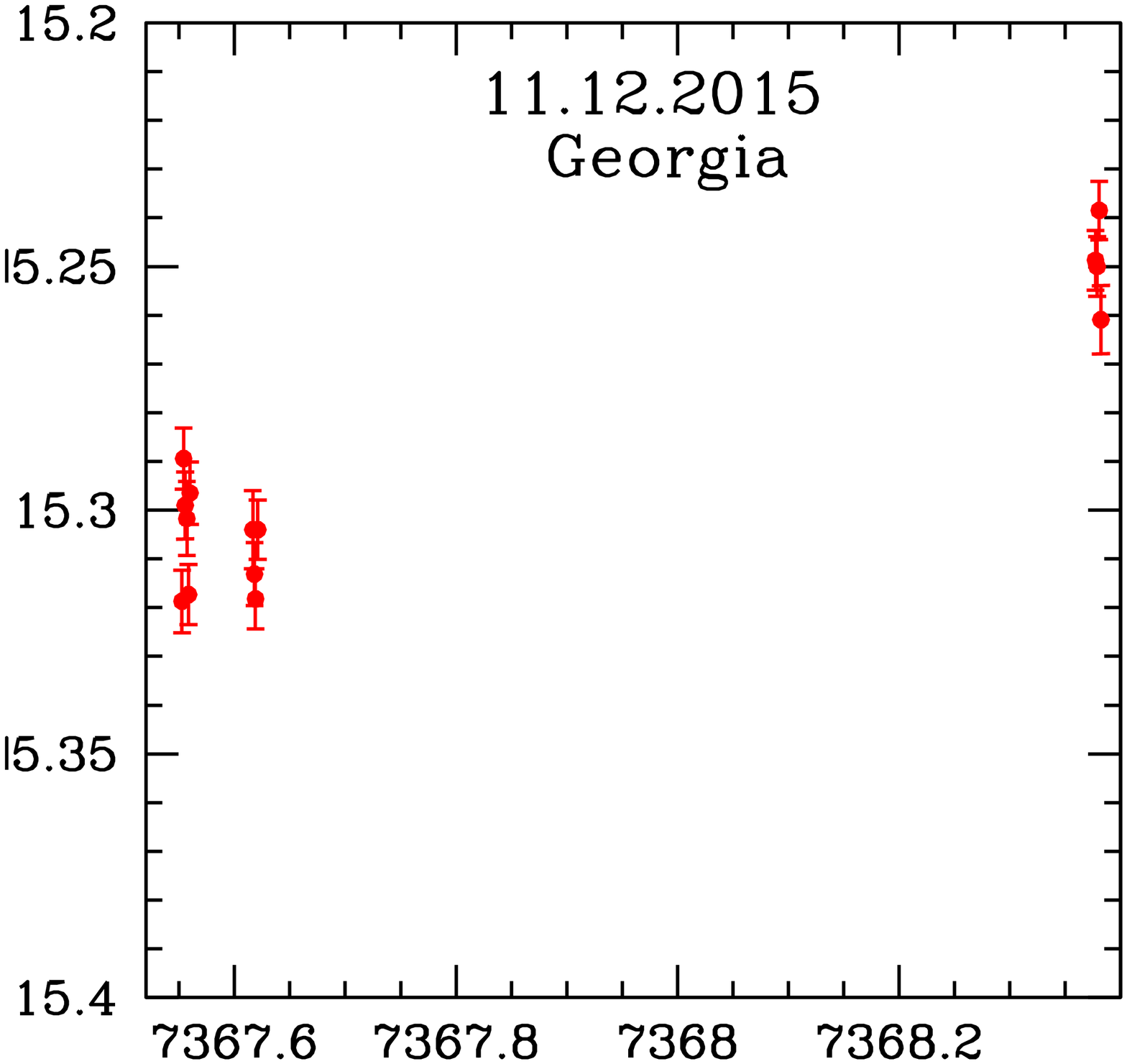,height=2.0in,width=2.0in,angle=0}
\epsfig{figure= 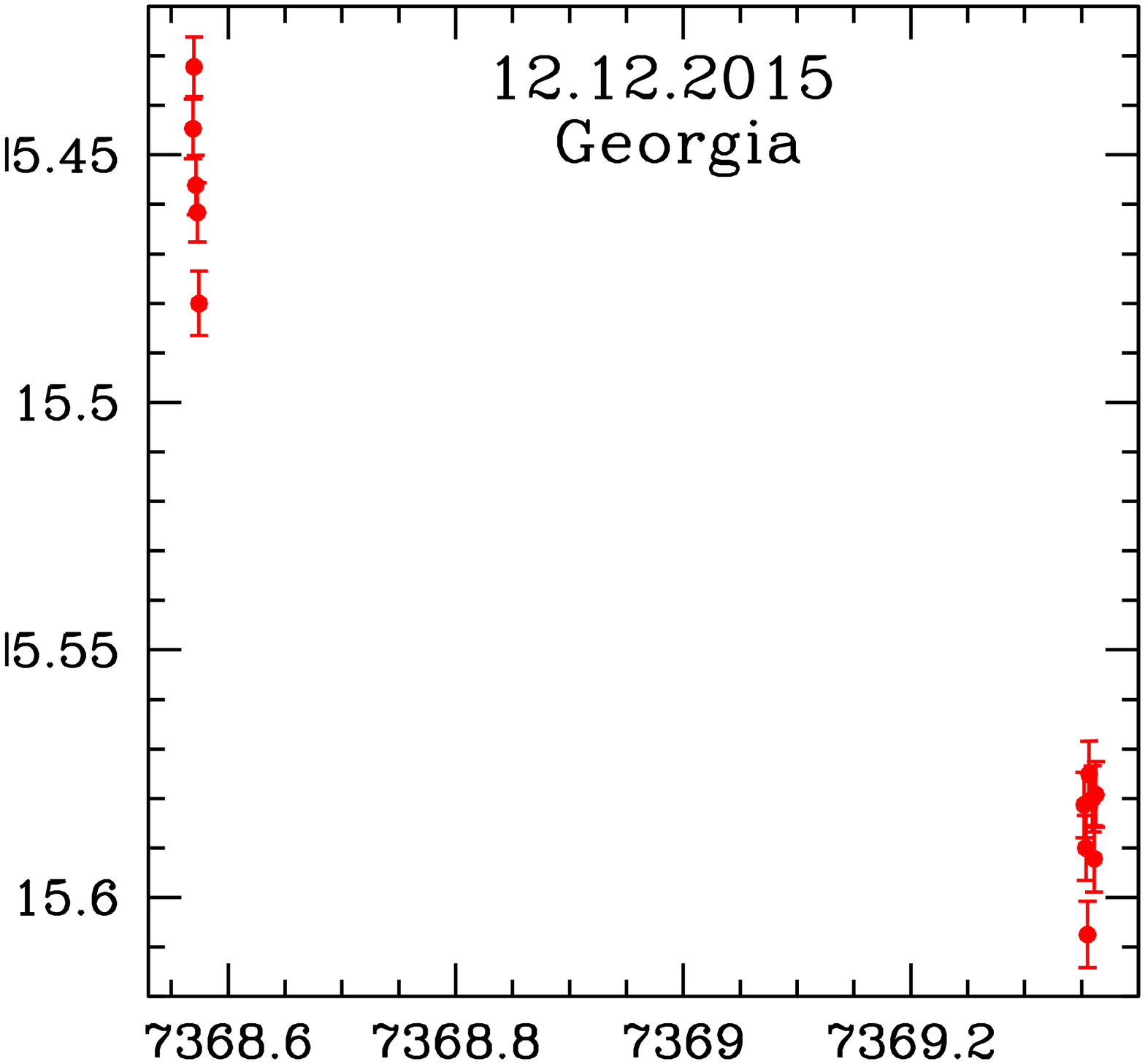,height=2.0in,width=2.0in,angle=0}
\epsfig{figure= 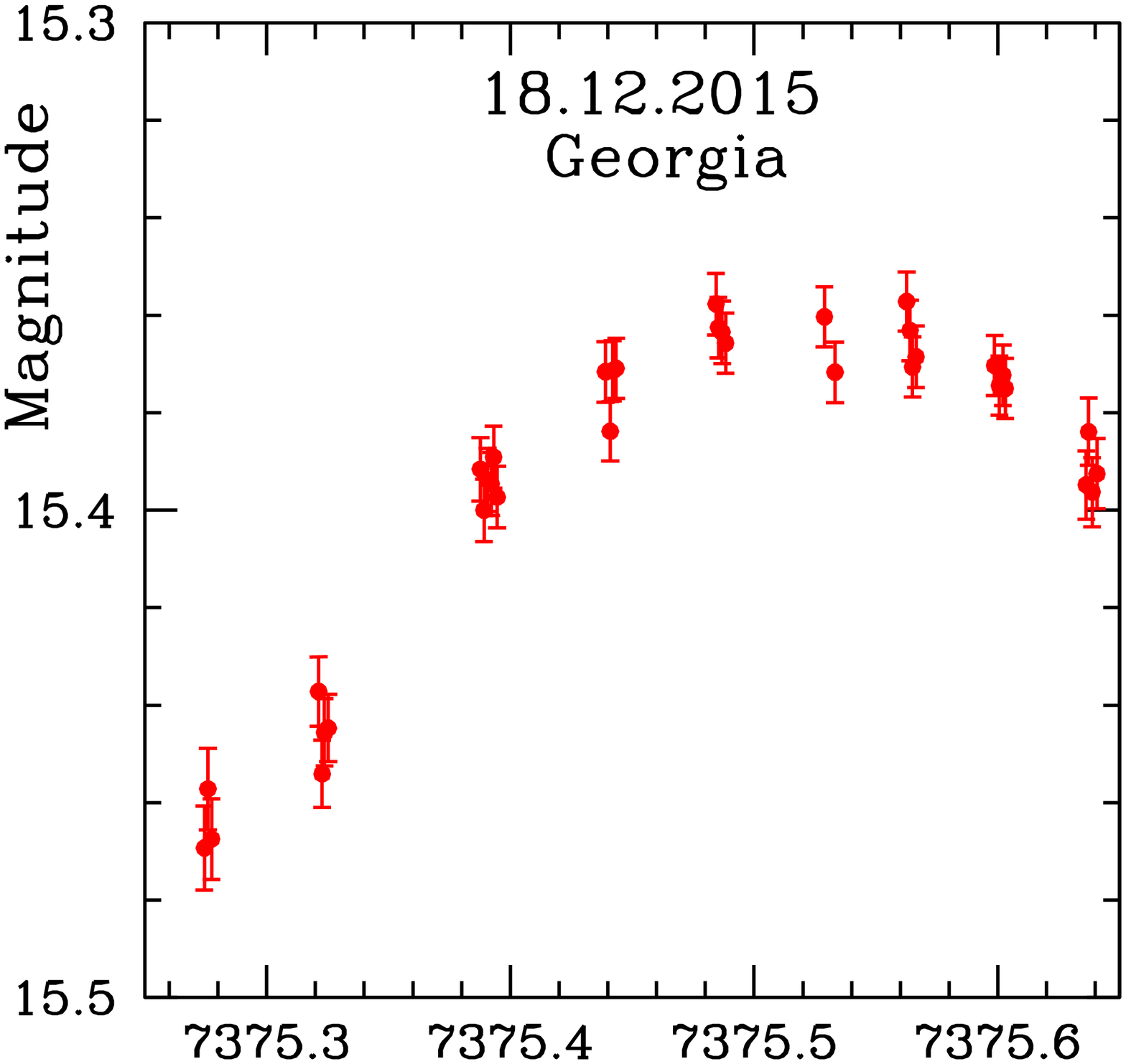,height=2.0in,width=2.0in,angle=0}
\epsfig{figure= 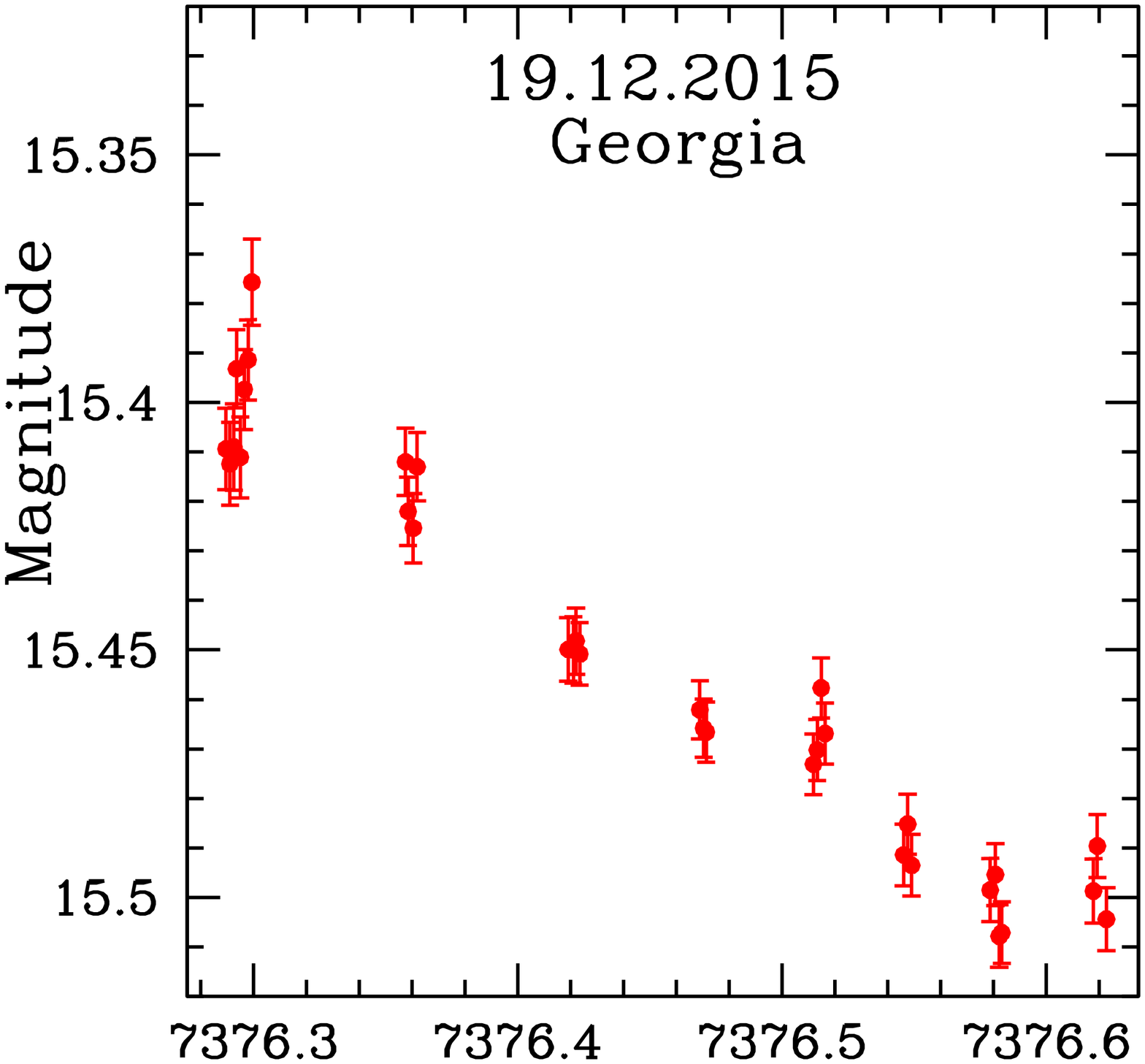,height=2.0in,width=2.0in,angle=0}
\epsfig{figure= 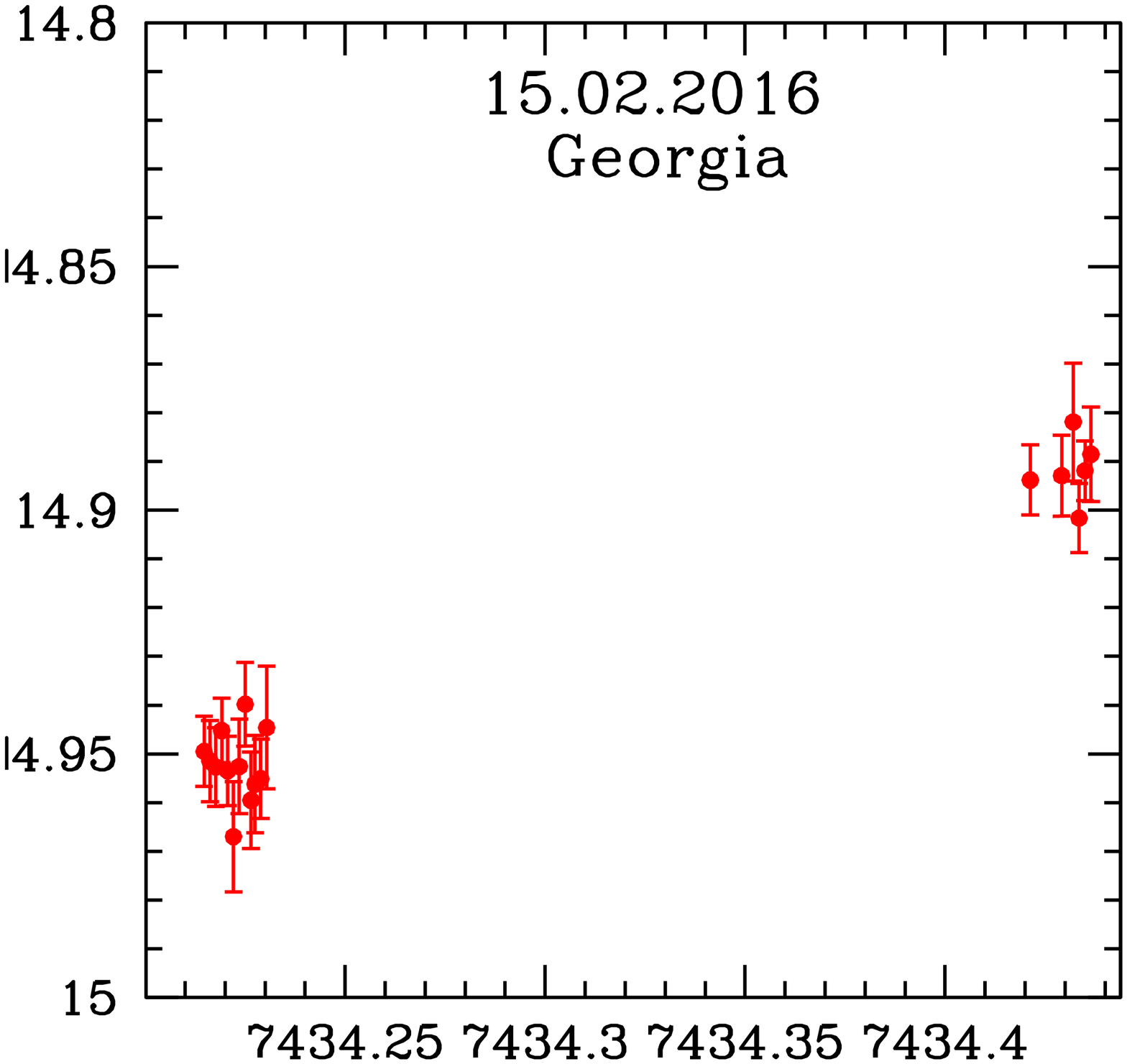,height=2.0in,width=2.0in,angle=0}
\epsfig{figure= 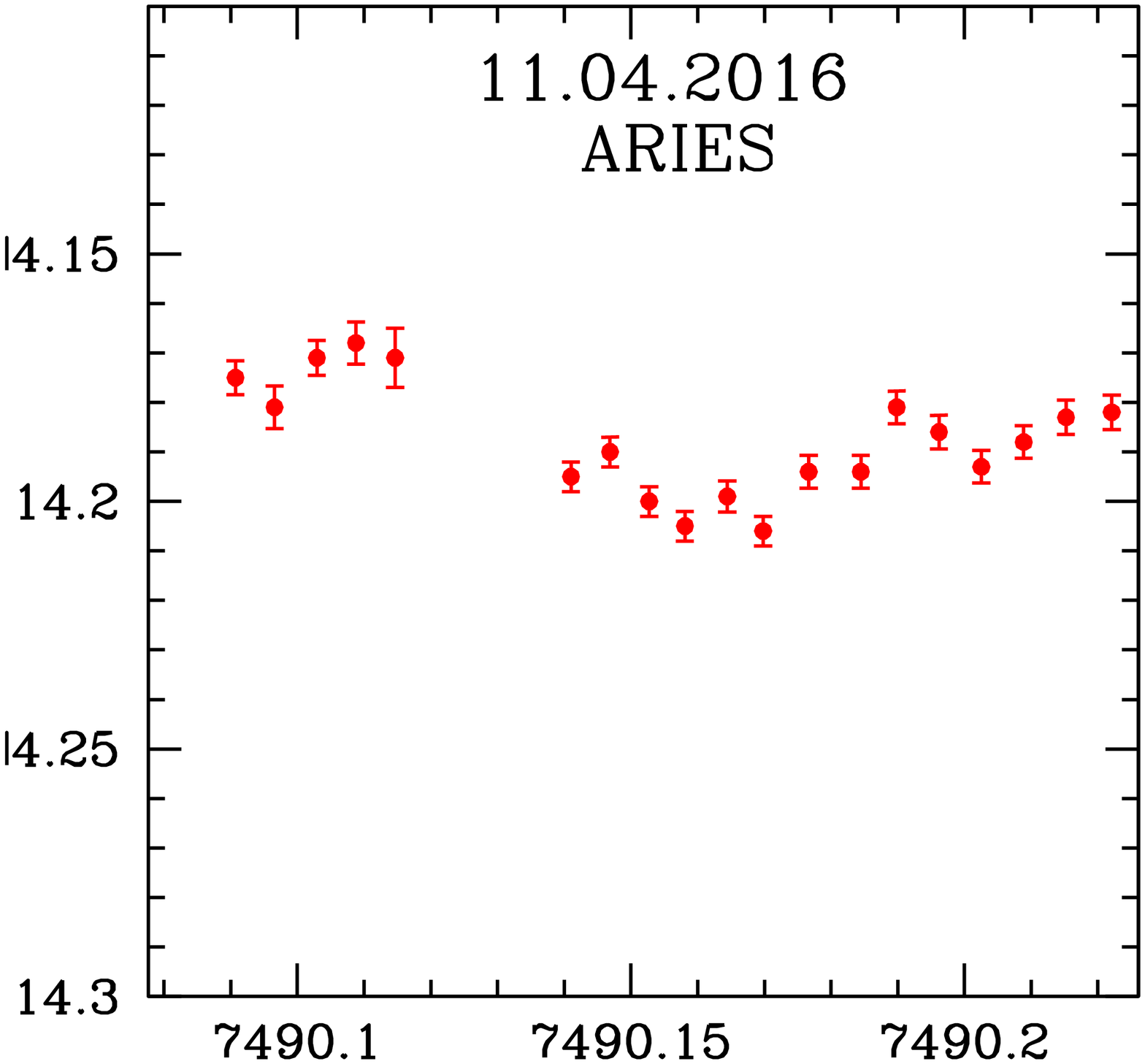,height=2.0in,width=2.0in,angle=0}
\epsfig{figure= 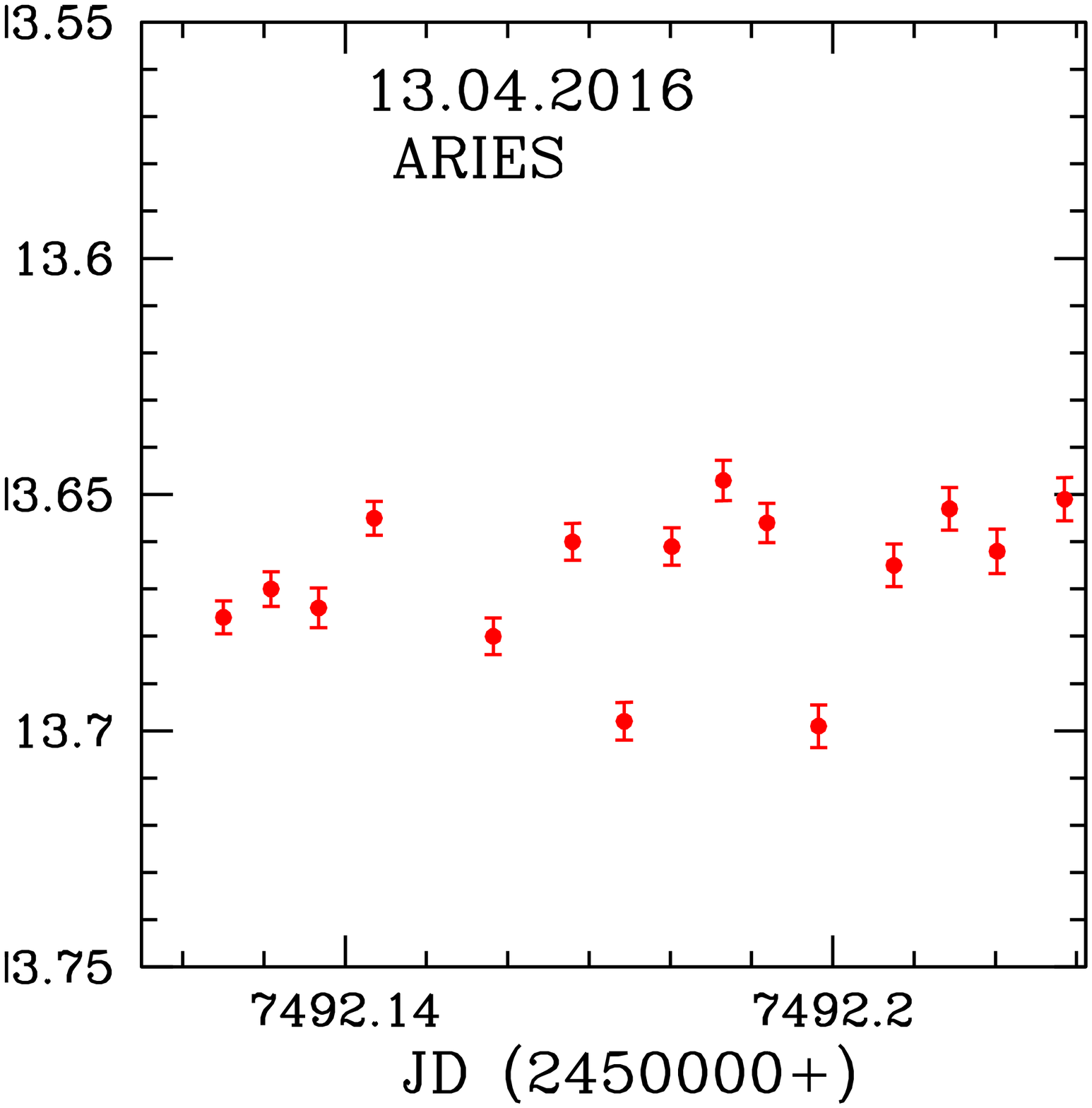,height=2.0in,width=2.0in,angle=0}
\epsfig{figure= 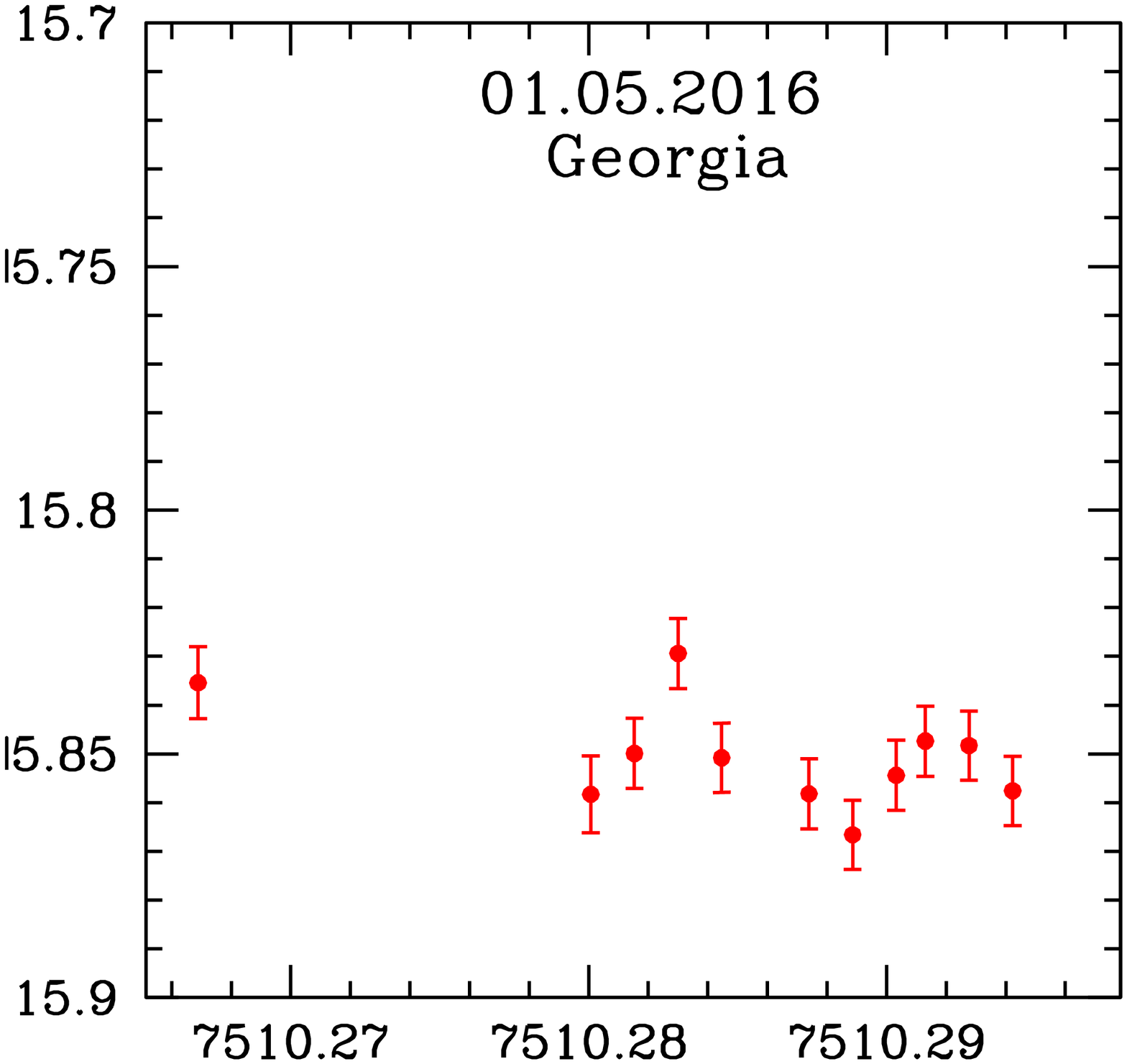,height=2.0in,width=2.0in,angle=0}
  \caption{Intraday light curves for OJ 287 in R filter. The X axis is JD and the Y axis is the magnitude in
  each plot, where observation dates are indicated in each plot.}
\label{LC_BL}
\end{figure*}

\begin{table*}
\caption{ Results of IDV observations of OJ 287. Column 1 is the date of observation, column 2 indicates
the band in which observations were taken, column 3 gives the number of data points in a particular band.
In the next two columns i.e.,  4 and 5, we have listed the results for the F-test and $\chi^{2}$-test,
respectively, followed by variability status in column 76; finally the intra-day variability amplitude is
given in column 7. }
\textwidth=7.0in
\textheight=10.0in
\noindent

\begin{tabular}{ccccccccc} \hline \nonumber

 Date       & Band   &N      & F-test  &$\chi^{2}$test  &   Variable    &A\% \\
            &        &       &$F_{1},F_{2},F,F_{c}(0.99),F_{c}(0.999)$ &$\chi^{2}_{1},
\chi^{2}_{2},\chi^{2}_{av}, \chi^{2}_{0.99}, \chi^{2}_{0.999}$  & & \\\hline 

 07.12.2015   & R     & 17  & 1.88, 2.41, 2.15, 3.37, 5.20    & 39.85, 69.42, 54.62, 32.00, 39.25    & PV  & -- \\
 11.12.2015   & R     & 14  & 16.72, 10.34, 13.53, 3.91, 6.41 & 270.62, 218.99, 244.81, 27.69, 34.53 & V & 7.95 \\
 12.12.2015   & R     & 12 & 176.08, 173.35, 174.72, 4.46, 7.76 & 2113.1, 2604.0, 2358.5, 24.72, 31.26 & V & 17.58 \\
 18.12.2015   & R     & 35 & 43.74, 45.30, 44.52, 2.26, 2.98 & 1381.0, 1880.0, 1631.0, 56.06, 65.25 & V & 11.16 \\
 19.12.2015   & R     & 33  & 54.38, 61.83, 58.10 2.32, 3.09  & 1757.0, 2592.0, 2174.0, 53.49, 62.49 & V & 13.16 \\
 15.02.2016   & R     & 12 & 0.81, 0.53, 0.67, 4.46, 7.76    & 12.59, 9.89, 11.24, 24.72, 31.26     & NV  & -- \\ 
 11.04.2016   & V     & 22  & 1.57, 0.91, 1.24, 2.86, 4.13    & 22775, 89536, 56156, 38.93, 46.80    & PV  & -- \\
              & R     & 19  & 2.61, 4.07, 3.34, 3.13, 4.68    & 67.24, 209.37, 138.30, 34.80, 42.31  & V  & 3.77 \\       
 12.04.2016   & V     & 24  & 0.04, 1.07, 0.56, 2.72, 3.85    & 4.89, 2.13, 3.51, 41.64, 49.73       & NV  & -- \\
              & R     & 24  & 2.00, 2.94, 2.47, 2.72, 3.85    & 131.08, 75.49, 103.29, 41.64, 49.73  & NV  & -- \\              
 13.04.2016   & V     & 18  & 1.22, 1.42, 1.32, 3.24, 4.92    & 0.03, 0.03, 0.85, 33.41, 40.79       & NV  & -- \\
              & R     & 17 & 1.00, 0.001, 0.50, 3.37, 5.20  & 1.66, 1.66, 0.85, 32.00, 39.25       & NV  & -- \\               
 01.05.2016   & R     & 11  & 11.18, 10.11, 10.65, 4.85, 8.75 & 102.20, 111.09, 106.65, 23.21, 29.59 & V & 7.67 \\ 
 
\hline
\end{tabular} \\
\noindent
V: Variable, PV: probable variable, NV: Non-Variable     \\
\end{table*}

\begin{figure*}
\epsfig{figure= 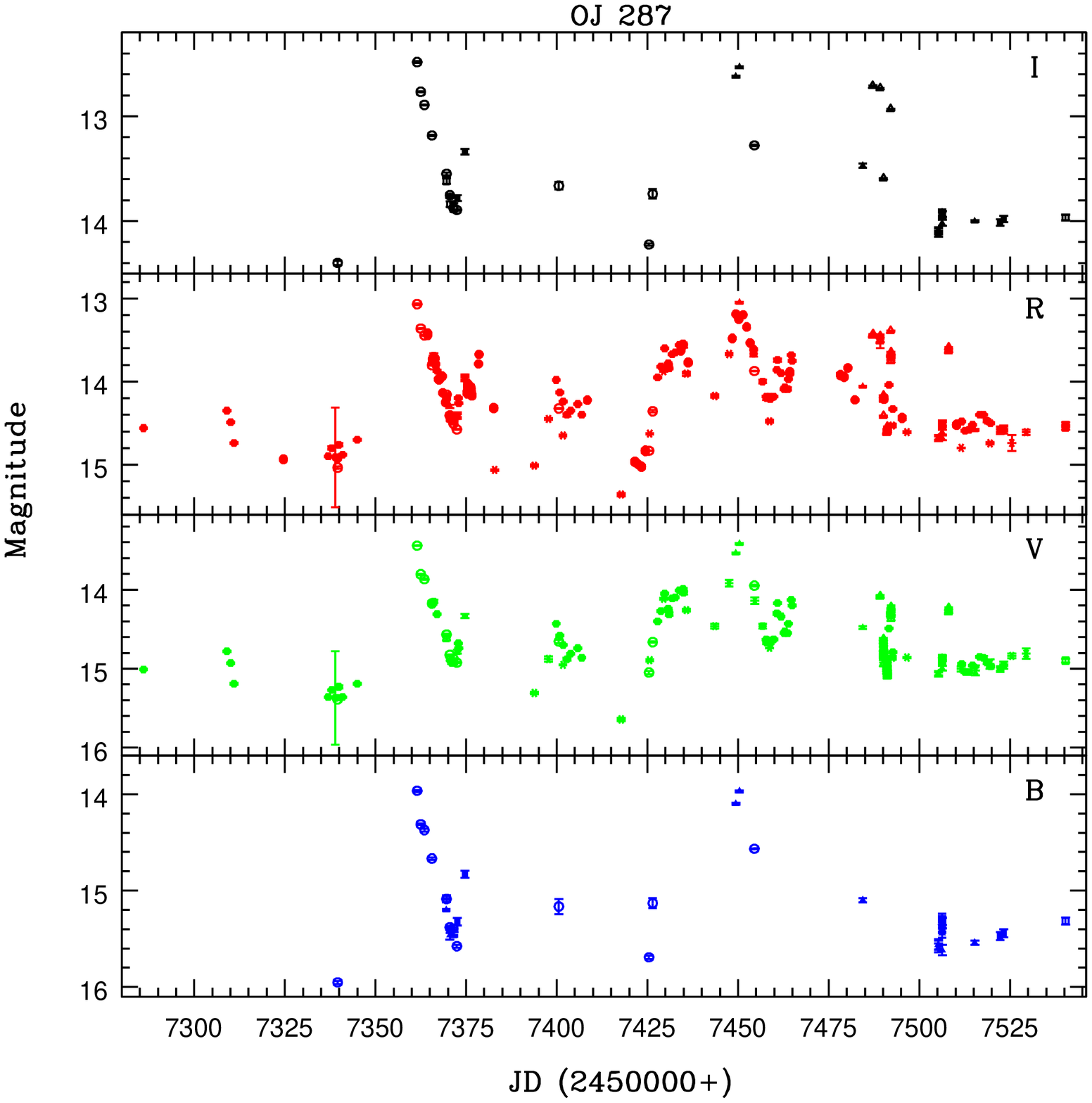,height=6.2in,width=6.2in,angle=0}
\caption{Optical variability light curve of OJ 287. Filled hexagon represent data from Arizona telescopes (A);
open triangle for the data from 1.04m India telescope (B); open and filled square for the data from 2m (C)
and 50/70 cm (D) telescopes Rozen, Bulgaria, respectively; filled triangle for the data from 60cm (E)
telescope Belogradchik, Bulgaria; open circle for the data from 60cm (F) telescope, Serbia; filled circle
for the data from 70cm (G) telescope, Georgia; and astrix for the data from 1.5m (H) telescope, Kanata, Japan.}
\end{figure*}

\section{Analysis Techniques}

\subsection{Variability detection criterion}

To test for the presence blazar variability on intraday timescales we have used two standard statistics, the
$F$ and $\chi^{2}$ tests (e.g., de Diego 2010).  As 
pointed out by Romero et al.\ (2002), inappropriate selection of standard stars could lead to spurious variability
detection. Thus we employed the two non-variable standard stars having magnitudes closest to OJ 287's magnitude. 
For calibration of the blazar magnitude we used the standard star having colour closer to the source.

\subsubsection{F$-$Test}

We  quantify the blazar variability with the commonly used F-test (de Diego 2010) which is defined as

\begin{equation}
 \label{eq.ftest}
 F_1=\frac{Var(BL-Star A)}{Var(BL-Star B)}, \nonumber \\ 
 F_2=\frac{Var(BL-Star A)}{ Var(Star A-Star B)}.
\end{equation}
Here (BL $-$ Star A), (BL $-$ Star B), and (Star A $-$ Star B) are the differential instrumental magnitudes
of blazar and Star A, blazar and Star B, and Star A and Star B, respectively, while Var(BL $-$ Star A),
Var(BL $-$ Star B), and Var(Star A $-$ Star B) are the variances of differential instrumental magnitudes.

An average of $F_1$ and $F_2$ gives the F value which is then compared with the critical $F^{(\alpha)}_{\nu_{bl},\nu_*}$ 
value where $\alpha$ gives the significance level set for the test while $\nu_{bl}$ and $\nu_*$ signifies the number of 
degrees of freedom and is calculated as (N $-$ 1) with N being the number of measurements. We have
carried out the F test for $\alpha$ values 0.999 and 0.99 percent which correspond to essentially $3 \sigma$ and $2.6 \sigma$
detections of IDV, respectively. The higher the $\alpha$ values, the more reliable is the result. When the F value is
greater than the critical value, the null hypothesis (no variability) is discarded. We consider LCs  to be
variable if $F > F_c(0.99)$.  

\subsubsection{$\chi^{2}-$Test}

To further quantify the variability of the blazar we also used a $\chi^{2}$-test. The $\chi^{2}$ statistic
is defined as (e.g., Agarwal \& Gupta 2015). 

\begin{equation}
\chi^2 = \sum_{i=1}^N \frac{(V_i - \overline{V})^2}{\sigma_i^2},
\end{equation}
where, $\overline{V}$ is the mean magnitude, and the $i$th observation gives a magnitude $V_i$ with
a corresponding standard error $\sigma_i$ which is due to photon noise from the source and sky, CCD
read-out plus other non-systematic error sources. Exact calculation of these errors by the IRAF reduction
package is impractical. Theoretical errors have been found to be larger than the real errors by a factor
of 1.3 $-$ 1.75 (e.g., Gopal-Krishna et al.\ 2003). Thus, the errors obtained after data analysis should
be multiplied by this factor, which was 1.5, to get  better estimates of the real photometric errors. The average of
$\chi^{2}$ values is then compared with the critical value $\chi_{\alpha,\nu}^2$ where $\alpha$ is again the
significance level and $\nu = N - 1$ is the number of degrees of freedom. Values of
$\chi^2 > \chi_{\alpha,\nu}^2$ imply the presence of variability.  We list sources as V only if they
satisfy both the F-test and the $\chi_2$-test for $\alpha = 0.99$.

\subsection{Percentage amplitude variation}

The percentage of magnitude and colour variations on intraday or longer time scales are calculated
using the variability amplitude parameter $A$, introduced by Heidt \& Wagner (1996), and defined as
\begin{eqnarray}
A = 100\times \sqrt{{(A_{max}-A_{min}})^2 - 2\sigma^2}(\%) .
\end{eqnarray}
Here $A_{max}$ and $A_{min}$ are the maximum and minimum values in the calibrated magnitude and
colour of LCs of the blazar, and $\sigma$ is the average measurement error.

\section{Results}

\subsection{Intraday flux and colour variability}

In searching for IDV, we used data from only those nights for which a minimum of 10 image frames of
the blazar were taken in any specific optical passband. Using this criterion, sufficient observations of the blazar OJ 287
 were carried out in the R passband on 10 nights and in the V passband on 3 nights 
between 2015 December and 2016 May. The differential light curves of the blazar were extracted by
following the data reduction procedure discussed above along with the differential instrumental magnitudes
of the two best standard stars (e.g. Gupta et al.\ 2008) used for comparison (Star A $-$ Star B). The 
calibrated derived LCs of the blazar are displayed in Fig.\ 1. V band LCs are not plotted due to their having
worse data
quality. 

In order to investigate the presence or absence of variability during above nights we
have followed the statistical analysis techniques described in section 3 with results summarized
in Table 3. The blazar is marked as variable (V) if the variability conditions for both tests mentioned
in Section 3 are satisfied; it is marked probably variable (PV) if conditions for one of the tests are
followed, otherwise it is marked non-variable (NV). According to these conditions the target was found to be 
variable on five nights, possibly variable on one night and non-variable on four of the nights in the R
passband. The blazar did not show any certain IDV in the V passband on any of the three nights it
was observed, though it was probably variable on the one when the R-band was variable. The blazar was in a 
generally active
state during our observation, as is most evident from the plots of 2015 December 11, 12 18, 19 and 2016 April 11 and 1 May  
in Figure 1 when clear variability patterns were seen. We have not displayed 2016 April 12 light curve in Figure 1 due
to poor quality data.

\subsection{Short and long term flux and colour variation}

We examined the short and long term flux and colour variability of the blazar. The overall LCs of OJ 287 are displayed
in Figure 2 giving our multi-band optical observations during September 2015 -- May 2016 for 103 observing
nights. The BVRI passband LCs presented in the Figure 2 give clear evidence of large amplitude flux
variability in all passbands on the timescale of days to weeks to months which gives clear evidence
of flux variability on short and long timescales.
  
During our observation we found the blazar has varied from $\sim$ 14 -- 16 mag in B band, $\sim$ 13.4 -- 15.4 mag
in V band, $\sim$ 13 to 15.4 mag in R band, and $\sim$ 12.5 to 14.4 mag in I band. Taking into account the noise, 
the variability amplitude
on long term basis in B band was calculated as 199\%, in V filter it was 222\%, in R passband 231\% while in
I band it was calculated to be 192\%. We also examined the (B-I), (V-R), (B-V), and (R-I)
colour indices on short and long timescale to search for  colour variability, and the results are plotted in Figure 3. 
While there were some modest apparent colour variations,  we did not find any
significant colour variability over a long term basis.

\subsection{Colour-magnitude relationship}

\begin{table}
\caption{Color-magnitude dependencies and colour-magnitude correlation coefficients on short timescales.}
\begin{center}
\begin{tabular}{ccccc} \hline

Color Indices     &  $m_1^a$  &  $c_1^a$  &   $r_1^a$  & $p_1^a$    \\ \hline
(B-I)          &  ~~~0.029 & ~~~1.565 & ~~~0.191  & 0.331  \\
(R-I)          &  $-$0.009 & ~~~1.637 & $-$0.103 & 0.588  \\
(B-V)          &  $-$0.012 & ~~~1.194 & $-$0.105 & 0.594 \\
(V-R)          &  ~~~0.016 & $-$0.736 & ~~~0.089 & 0.370 \\\hline
\end{tabular}
\end{center}
\noindent
$^a$ $m_1 =$ slope and $c_1 =$ intercept of CI against V; \\
$r_1 =$ Pearson coefficient; $p_1 =$ null hypothesis probability \\
\end{table}

Since flux variations are often associated with changes in spectral shapes, we also studied the correlations
between the colour indices and flux variations. Here we have considered  whether variations in the
(B$-$V), (V$-$R), (R$-$I) and (B$-$I) colour indices of the blazar change with respect to variations in its 
brightness in the V passband on short and long term bases. These colour and magnitude plots of OJ 287 are displayed
in Fig. 4. 
We have calculated the best linear fit as shown by the straight lines in the Figure 4 for each colour index, $Y$,
against magnitude, $V$: $Y = mV + c$. Fitted values for the slopes of the curves, $m$, and the constants, $c$,
are listed in Table 4. We have also listed the linear Pearson correlation coefficients, $r$, and the
corresponding null hypothesis probability values, $p$. Positive slope between the colour index and apparent
magnitude of the blazar implies a positive correlation which further means that the source tends to be bluer
when it is brighter (BWB) or redder when it dimmer while a negative slope suggests an opposite correlation 
i.e. the source follows a redder when brighter (RWB) behaviour.  Quite a few blazars have been found
to show BWB behaviour (e.g., Raiteri et al.\ 2001; Villata et al.\ 2002; Papadakis et al.\ 2003; Agarwal \& Gupta 2015).
We find no significant negative or positive correlation between 
the V band magnitude and colour indices during these observations of OJ 287. 
Earlier studies (e.g., Raiteri et al.\ 2003; Wu et al.\ 2005; Stalin et al.\ 2006; Agarwal et al.\ 2016;
and references therein) also found at the most weak or null evidence of spectral changes with the source
brightness on short and long timescales. No spectral change with magnitude variations have been also been
reported by many authors in several other cases (e.g. Ghosh et al.\ 2000; B{\"o}ttcher et al.\ 2007; Poon et al.\ 2009;
and references therein).

\begin{figure}
\epsfig{figure= 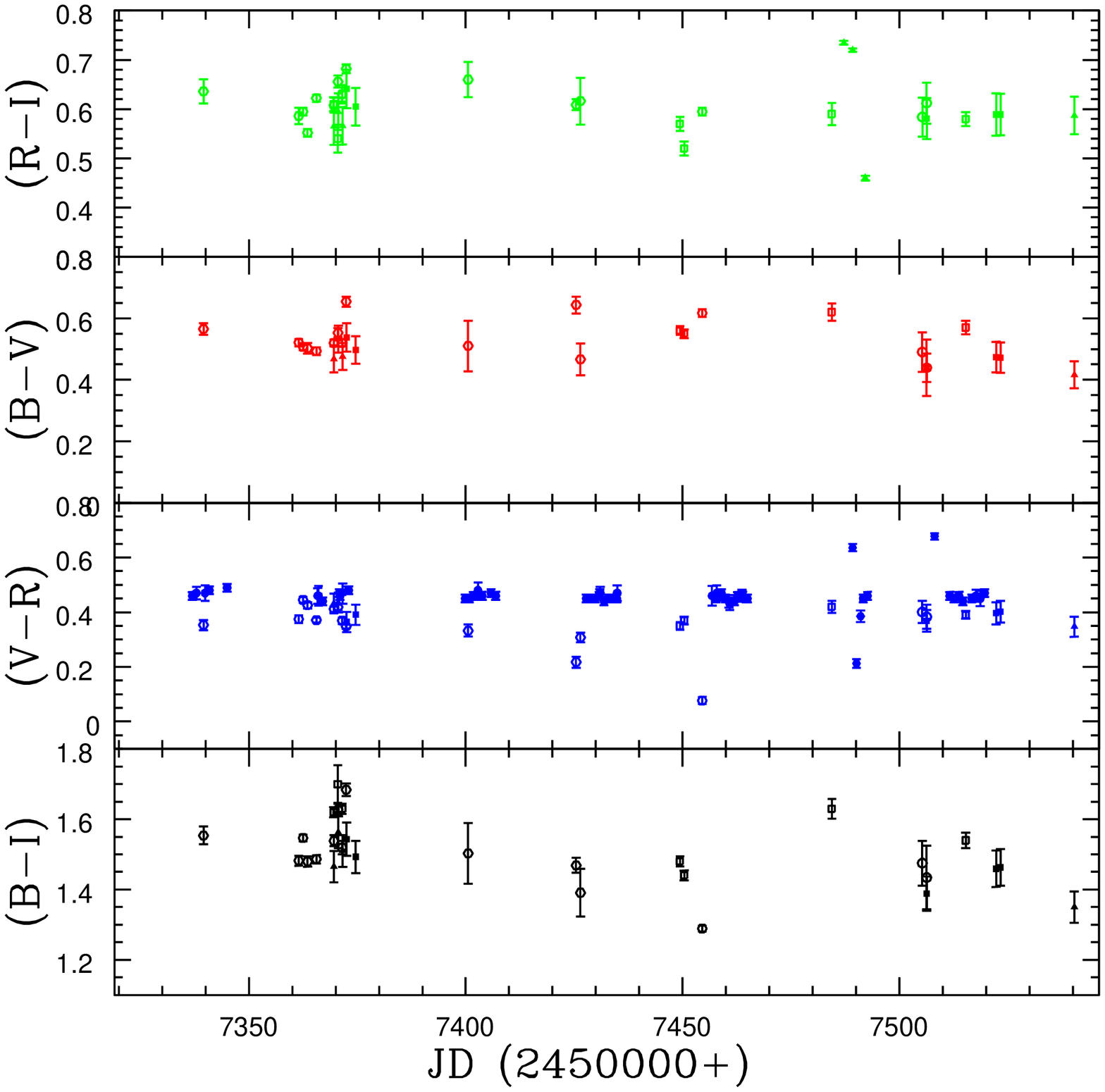,height=3.2in,width=3.2in,angle=0}
\caption{Optical colour variability light curves of OJ 287 covering the entire monitoring period.}
\end{figure}

\begin{figure}
\epsfig{figure= 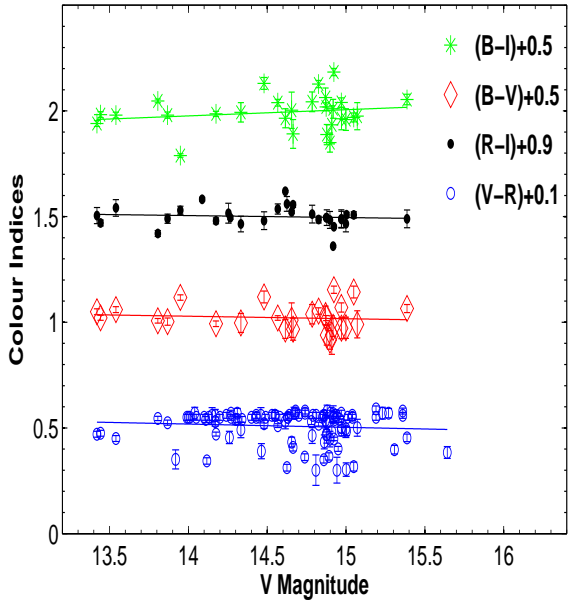,height=3.4in,width=3.4in,angle=0}
\caption{Optical colour-magnitude plots of OJ 287 during our monitoring.}
\end{figure}

\begin{figure}
\epsfig{figure= 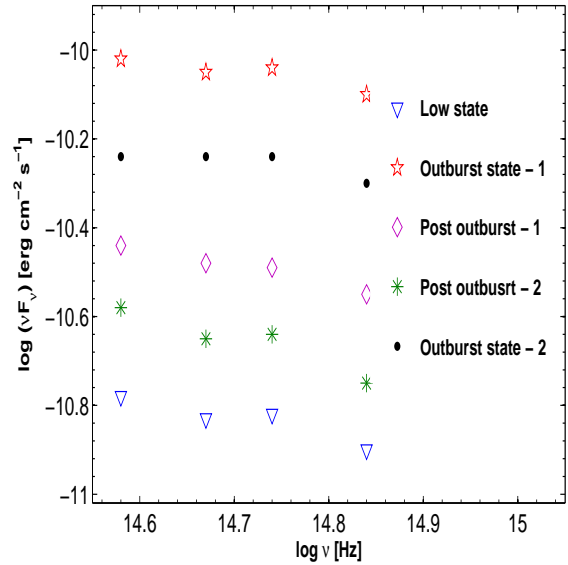,height=3.2in,width=3.2in,angle=0}
\caption{Optical SED of OJ 287 during different states. An offset of $-$0.2 for  outburst state-2 is used
to avoid its overlapping the SED of outburst state-1.}
\end{figure}

\begin{figure}
\epsfig{figure= 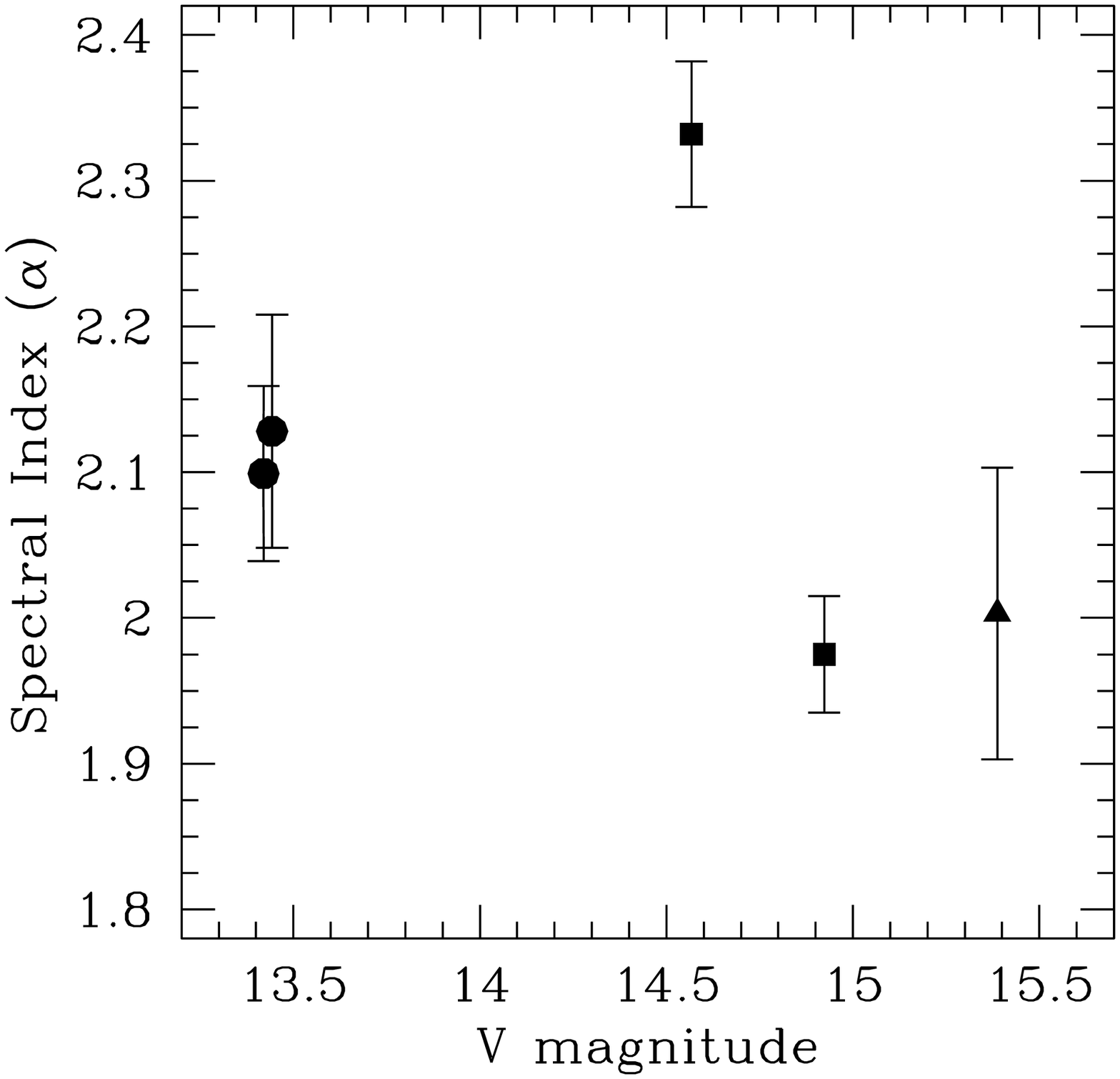,height=3.2in,width=3.2in,angle=0}
\caption{Variation of optical spectral index with V magnitude during outburst states (filled circles),
post-outburst states (filled squares) and faint state (filled triangle) of OJ 287.}
\end{figure}

\subsection{Optical Spectral Energy Distribution}

The SED is a simple but useful tool to diagnose the dominant emission components. Flux changes in blazars are often directly
linked to the spectral changes, thereby making SED studies an important tool to learn 
details about the emission region, and can sometimes put tight constraints on the physical parameters such as the magnetic field,
or energies of the relativistic particles in the Doppler boosted jet.
To understand spectral variability in optical range, we used quasi-simultaneous B, V, R, and I data
points at five different states of the source. The outburst state 1 is is taken as  that epoch when the
source attained the brightest value in all four bands which lies near JD 2457361.5 (2015 Dec 5).  We note that we do not
have observations during the preceding two weeks so from our data the actual peak may have been even greater;
however, the monitoring by Valtonen et al.\ (2016) had better coverage during the large flare and showed that it rose
during that period and that the maximum brightness was reached on 2015 Dec 5.  As is
evident from the STV LC another very comparable outburst was observed around JD 2457450.4 (2016 March 2) which we call
as outburst state 2 and which is covered well by our observations but not covered by Valtonen et al.\ (2016). 
To quantify the spectral variability changes in the source during its faint state,
we generated the SED around JD 2457339.6 (2015 Nov 13) when the source magnitude reached its maximum value.
We also considered two stages in the post outburst state on JDs 2457369.6 (2015 Dec 13) and 2457372.4
(2015 Dec 15).

We have de-reddened the calibrated magnitudes of OJ 287 by subtracting Galactic absorption values of
A$_{B}$ = 0.102 mag, A$_{V}$ = 0.077 mag, A$_{R}$ = 0.061 mag, and A$_{I}$ =  0.042 mag
(Cardelli et al.\ 1989; Bessell et al.\  1998). We took single data points in B, V, R, and I passbands,
corresponding to the JDs for each of the above states and thus generated optical SEDs at five epochs using our 
quasi-simultaneous B, V, R, and I observations of OJ 287. These  SEDs are displayed in Figure 5.
The faintest SED was observed for JD 2457339.6 while the brightest one was observed for JD 2457361.5.
Significant spectral changes are clearly evident from the figure as the source varied from its lowest
brightness state to the highest one.  The presence of a bump around the V-B bands is most visible in the SEDs corresponding to
the post outburst state 2 and the Low state, as displayed in Fig.\ 5. The little blue bump
might be attributed to the emission lines due to BLR region in the rest wavelength range $\sim$ 200 -- 400 nm
(Raiteri et al.\ 2007), but in this case thermal emission from the accretion disc gas impacted by the
secondary black hole is the most likely explanation (Valtonen et al.\ 2016). 
We also calculated the average spectral indices for above five epochs using 
(Wierzcholska et al.\ 2015),
\begin{equation}
\langle \alpha_{VR} \rangle = {0.4\, \langle V-R \rangle \over \log(\nu_V / \nu_R)} \, ,
\end{equation}
where $\nu_V$ and $\nu_R$  are effective frequencies of the respective bands (Bessell, Castelli, 
\& Plez 1998).
The spectral index varied between 1.97 to 2.33 as is visible in Figure 6 where we have plotted spectral
indices at various stages of the source vs the V band magnitude. The $\alpha_{VR}$ value for the low
brightness state of the source was about 2.00 while for the outburst state 1 is $\sim$ 2.13. The higher
values of spectral indices indicate synchrotron dominated emission due to relativistic Doppler boosted jet
emission (e.g. Kushwaha et al. 2013). Simultaneous multi-wavelength observations would be helpful in
having detailed understanding of the spectral evolution of the target during its double outburst state.

\subsection{Flaring and outbursts}

Over the roughly nine months (September 2015 -- May 2016) of our intense optical observing campaign of the
blazar OJ 287, this source showed several unprecedented large amplitude flares seen best in
the most densely observed  R band data plotted in the second panel from the top in Figure 2. In our data there is
a large gap during $\sim$ JD 2457340 to JD 2457360 but that period was intensely monitored by Valtonen et al.\ (2016).
The source went into an optical outburst at JD 2457342.5$\pm$2.5 (Valtonen et al.\ 2016) and reached
the brightest value at R $\sim$ 13.05 mag on JD 2457361.5 (i.e., in $\sim$ 20 days) and then decayed to R $\sim$
14.65 mag at about JD 2457372 (i.e., in $\sim$ 10 days). Then another flare started at about JD 2457372 at
R $\sim$ 14.65 which we observed as it reached R $\sim$ 13.6 mag at about JD 2457372. A third significant
flare started at about JD 2457423 when R $\sim$ 15 mag and reached R $\sim$ 13.4 in about 12 days or
JD 2457435. The fourth large flare is  one that could be considered a second  outburst, as it reached R $\sim$ 13.05 on 
about JD 2457450, when it had essentially the same brightness of the first outburst at maximum; however, in the binary black hole model, all of these flares would be considered to be part of the first outburst, with the second outburst expected to lag by at least a year. Another interesting feature 
in this overall outburst is that the largest flares seen during our monitoring (one and four)  are separated by $\sim$ 3 months.     

\section{Discussion}

Blazars, being dominated by jet emissions, display variability on a very wide range of timescales. The reason behind this
blazar variability could be either intrinsic to the jet or due to changes in overall jet power. Shock-in-jet models
(e.g., Marscher \& Gear 1985; Spada et al.\ 2001; Graff et al.\ 2008; Joshi \& B{\"o}ttcher 2011) 
appear to explain the longer term variations quite well for most blazars.  Changes in any of the physical parameters like velocity, 
electron density or magnetic field can trigger the emergence of a new shock which leads to high amplitude
flares when propagating along the inhomogeneous medium of the relativistic jet.  The moving shock causes
acceleration of electrons to even TeV energies thus leading to  radiation changes in all EM bands.
Precession, helical structures or various other
geometrical effects related to the Doppler boosted relativistic jet could lead to significant changes on
STV timescales though they are probably more relevant for LTV ones (e.g. Camenzind \& Krokenberger 1992;  Pollack et al.\ 2016).
Valtonen \& Pihajoki (2013) have applied such a helical jet model to OJ 287.  Some blazar
variability can be also be extrinsic in nature.  Such extrinsic mechanisms include refractive interstellar 
scintillations which, however, is only relevant for low frequency radio observations.
Another extrinsic mechanism is gravitational microlensing which is applicable over the entire EM spectrum
but is only relevant for lensed sources. Since we are reporting results based on multi-band
optical observations of the blazar OJ 287 which is not a lensed system,  extrinsic causes of variability
are ruled out. 
Turbulence behind the shocks in a jet provide a good way of understanding much of the fastest variability seen from
blazars (Marscher 2014; Calafut \& Wiita 2015; Pollack et al.\ 2016).
Blazar variability on IDV timescale could be absent if there is no change in the speed or the direction
of the propagating shock along the observer's LOS. 
IDV and STV in quasars and perhaps in blazars during their low state can be attributed to instabilities in the accretion disc
such as hot spots (e.g., Chakrabarti \& Wiita, 1993; Mangalam \& Wiita, 1993). Disturbances on or above the
accretion disks are advected into the relativistic jet and thus, are Doppler amplified, which in turn
might also provide seed fluctuations.   

In the (so far) unique case of OJ 287 the strong flares occurring roughly every 12 years are
almost certainly produced by  large perturbations induced in the accretion disc around the primary supermassive
black hole as the secondary black hole smashes through the disc (e.g.\ Lehto \& Valtonen 1996;  Valtonen et al.\ 2016).  
Because of the only quasi-Keplerian nature of the binary black hole orbits in general
relativity and the impact of gravitational radiation on the orbits, the impact outbursts and their associated electromagnetic 
radiation will not be observed in a strictly periodic manner (Valtonen, Ciprini \& Lehto 2012). In this model, thermal outbursts are 
generated by expanding
bubbles of hot gas that have been shocked and pulled out of the accretion disc around the primary black hole. 
It was predicted that OJ 287 should show a major outburst in 2015
December (Valtonen et al.\ 2011). In our work, along with confirmation of this predicted major outburst during late 2015 also reported by
Valtonen et al.\ (2016), we have found a comparably strong flare with a clear signature of a bump in the V and B bands  after just 90 days.   In the previous 2005--2007 cycle there was also a substantial event around 90 days after the primary outburst (Valtonen \& Ciprini 2012), though it was not as relatively strong as we have found now.  A possible explanation for this event is that it corresponds to a jet flare arising after this transfer time from the site of the impact to the base of the jet (Valtonen et al.\ 2006; Pihajoki et al.\ 2013).
In the binary black hole model it expected that another comparable outburst will occur one to two years after the primary one, to parallel the  gaps seen between the  events during the previous two cycles in 1994--1995 and 2005--2007, so OJ 287 should continued to be monitored whenever possible. 

Colour-magnitude (CM) correlations for blazars have been carried out by several authors on intraday
as well as on short term bases (e.g.\ Papadakis et al.\ 2003, Sasada et al.\ 2010, Bonning et al.\ 2012,
Gaur et al.\ 2012c, Agarwal \& Gupta 2015). Such CM studies can elucidate the reason behind aspects of
blazar variability
and also help us to explore the emitting region. Significant BWB/RWB trends on diverse timescales with 
different slope values are well known in blazars (Gaur et al.\ 2012c, Agarwal et al.\ 2016) and 
are sensibly interpreted as arising from superpositions of red and blue emission 
components. For BWB situations the blue component can be ascribed to synchrotron emission from the Doppler boosted
relativistic jet while the thermal emissions from the accretion disc could give the redder contribution. 
No clear BWB or RWB trends were found in these data. To be able to detect
weaker CM relationships, we need to obtain very dense and highly precise simultaneous multi-band observations.
These are needed because there are chances of non-negligible magnitude fluctuations occurring while changing from one filter to another during the 
quasi-simultaneous observations making it much more difficult to accurately  measure colour changes.

\section{Conclusions}

We carried out multi-band optical photometric observations of the blazar OJ 287 during 103 observing nights
between September 2015 -- May 2016 using 10 ground based optical telescopes located all around the globe.
We searched for flux variation on IDV, STV and LTV timescales, and for colour variations and any colour - magnitude
relationship in the blazar time series data on STV and LTV timescales. Our main conclusions are following:

$\bullet$ We observed OJ 287  intensely enough to perform IDV studies in R and V passbands on 11 and 4 observing nights,
respectively. We did not
find significant IDV on any of the nights observed in the V band (though it was probably variable on one of them)
while the blazar did show IDV in the R passband on 6 nights
and possible IDV on 1 additional night. 

$\bullet$ The blazar was in a highly active and variable state during our observing period, showing 
large amplitude flux variations on STV and LTV timescales. OJ 287  varied by $\sim$ 199\%, 222\%, 231\%
and 192\% in the B, V, R and I passbands respectively. 

$\bullet$ OJ 287 did not show any significant colour variations on STV and LTV timescales, nor
did the colour--magnitude relationship  indicate any significant spectral variations on those timescales.

$\bullet$ The blazar went through multiple flaring events during our observing run and there were
four strong flares.  The first flare occurred at the time predicted by the binary black hole model for OJ 287
(Valtonen \& Ciprini \ 2012).

$\bullet$ OJ 287 again showed a major flare separated by $\sim$ 90 days from the first one, and this last of the 
four flares we detected was exceptional in that it peaked  at nearly the same magnitude as did the first. 

\section*{ACKNOWLEDGMENTS}

We thank the referee for constructive comments and suggestions.
Data from the Steward Observatory spectropolarimetric monitoring project were used. This program is supported
by Fermi Guest Investigator grants NNX08AW56G, NNX09AU10G, NNX12AO93G, and NNX15AU81G.
ACG is partially supported by CAS President's International Fellowship Initiative (PIFI) (grant no. 2016VMB073).
HG is sponsored by the Chinese Academy of Sciences (CAS) Visiting Fellowship for Researchers from Developing
Countries, CAS Presidents International Fellowship Initiative (grant no. 2014FFJB0005), supported by the NSFC
Research Fund for International Young Scientists (grant no. 11450110398) and supported by a Special Financial
Grant from the China Postdoctoral Science Foundation (grant no. 2016T90393). MFG is supported by the 
National Science Foundation of China (grants 11473054 and U1531245) and by the Science and Technology Commission 
of Shanghai Municipality (grant 14ZR1447100). AS, ES, RB work was partially
supported by Scientific Research Fund of the Bulgarian Ministry of Education and Sciences under grant DO 02-137
(BIn-13/09). GD and OV gratefully acknowledge the observing grant support from the Institute of Astronomy and
Rozhen National Astronomical Observatory, Bulgaria Academy of Sciences, via bilateral joint research project
``Observations of ICRF radio-sources visible in optical domain". This work is a part of the project nos. 176011
176004, and 176021 supported by the Ministry of Education, Science and Technological Development of the Republic
of Serbia. The Abastumani Observatory team acknowledges financial support by the by Shota Rustaveli National
Science Foundation under contract FR/577/6-320/13.

\clearpage

\end{document}